\documentstyle[twoside,fleqn,espcrc2]{article}

\input epsf.sty
\newdimen\psfigsize
\def\psfigure#1 #2 #3 #4 #5{
    \begin{figure}[tbh]
    \vbox{
    \null\vskip-0.2in\hskip#2\epsfxsize=#1 \epsfbox[0 0 4096 4096]{#4}
    \vskip -0.3in
    \caption {#5 \label{#3}}
    \vskip 0.0truein plus0.2truein}
    \end{figure}
}
\def\psoddfigure#1 #2 #3 #4 #5 #6{
    \begin{figure}[tbh]
    \vbox{
    \null\vskip-0.2in\hskip#3\epsfxsize=#1 \epsfbox[0 0 4096 4096]{#5}
    \vskip -#1 \vskip #2 \vskip 10truept
    \vskip -0.2in
    \caption {#6 \label{#4}}
    \vskip 0.0truein plus0.2truein}
    \end{figure}
}

\def\BE{\begin{equation}}
\def\EE{\end{equation}}
\def\BEA{\begin{eqnarray}}
\def\EEA{\end{eqnarray}}

\def\rightpartial{{\overrightarrow\partial}}
\def\leftpartial{{\overleftarrow\partial}}
\def\diffpartial{\buildrel\leftrightarrow\over\partial}

\def\eps{\epsilon}
\def\pbp{\bar\psi\psi}

\def\psibar {\bar \psi}

\newcommand{\AmS}{{\protect\the\textfont2
  A\kern-.1667em\lower.5ex\hbox{M}\kern-.125emS}}

\title{Exotic hybrid mesons with light quarks}

\author{ Claude~Bernard,\hskip-0.03in
\address{{\vskip-0.10in{\hskip 0.07in Department of Physics, Washington University, St.~Louis, MO 63130, USA}}} 
Tom~Blum,\hskip-0.03in
\address{Department of Physics, Brookhaven National Lab, Upton, NY 11973, USA} 
Thomas~A.~DeGrand,\hskip-0.03in
\address{Physics Department, University of Colorado, Boulder, CO 80309, USA} 
Carleton~DeTar,\hskip-0.03in
\address{Physics Department, University of Utah, Salt Lake City, UT 84112, USA} 
Steven~Gottlieb,\hskip-0.03in
\address{Department of Physics, Indiana University, Bloomington, IN 47405, USA} 
Urs~M.~Heller,\hskip-0.03in
\address{SCRI, Florida State University, Tallahassee, FL 32306-4052, USA} 
Jim~Hetrick,\hskip-0.03in
\address{Department of Physics, University of Arizona, Tucson, AZ 85721, USA} 
Craig~McNeile,\hskip-0.03in$\,\null^{\rm d}$
Kari~Rummukainen,\hskip-0.03in$\,\null^{\rm e}$
Bob~Sugar,\hskip-0.03in
\address{Department of Physics, University of California, Santa Barbara, CA 93106, USA}
Doug~Toussaint$\,\null^{\rm g}$
\thanks{presented by Doug~Toussaint}
and Matt~Wingate$\,\null^{\rm c}$
} 
       
\begin{document}

\begin{abstract}
Hybrid mesons, made from a quark, an antiquark and gluons, can have quantum
numbers inaccessible to conventional quark-antiquark states.  Confirmation
of such states would give information on the role of ``dynamical'' color in
low energy QCD. We present preliminary results for hybrid meson
masses using light Wilson valence quarks.
\end{abstract}
\maketitle

\section{Introduction}

QCD, the dynamical theory of strong interactions, appears to predict
the existence
of hadrons beyond those in the simple quark model, namely glueballs and
hybrids.
Confirmation
of such states would give information on the role of ``dynamical'' color in
low energy QCD.
Hybrid mesons, made from a quark, an antiquark and one or more
gluons, can have quantum
numbers inaccessible to conventional quark-antiquark states.
Isospin triplet hybrid mesons with such ``exotic'' quantum numbers are
especially interesting because they cannot mix with conventional hadrons
or with glueballs.  Bound states of quarks and gluons are obviously
nonperturbative phenomena, and in principle lattice gauge theory is the
ideal method for calculating their properties, although in practice this
turns out to be difficult.

The earliest lattice calculations of hybrid mesons used heavy
quarks, where hybrid states appear as excitations of the gluonic
string\cite{MICHAELANDPERANTONIS,UKQCDOLD}.
Here we present preliminary results for hybrid
masses using lighter Wilson valence quarks, although not as light as the
physical up and down quarks.
The UKQCD collaboration
has similar results for hybrid mesons\cite{MICHAEL,UKQCD}.

\section{Hybrid Operators}

To make an operator which creates a hybrid meson, we
combine a quark, an antiquark and the color electric or magnetic field to form
a color singlet with the desired spin, parity and charge conjugation.
Because we do not include ``quark-line disconnected'' diagrams in our
propagator, all our meson propagators are isospin triplets.
The color electric and magnetic fields have $J^{PC} = 1^{--}$ and
$1^{+-}$ respectively. The spin, parity and charge conjugation from
the quark and antiquark are those of the available quark bilinears
listed here along with the corresponding mesons:
(The $0^{+-}$ bilinear, $\psibar \gamma_0
\psi$, is not expected to produce a particle since it is the charge
corresponding to a conserved current --- the baryon number.  We expect
$\int d^3 x \psibar \gamma_0\psi \, | 0 \rangle = 0$.)
\vskip 0.05in

\begin{tabular}{lll}
$0^{++}$ & ($\pbp$) & ($a_0$)\\ 
$0^{+-}$ & ($\psibar \gamma_0 \psi$) & (NONE) \\
$0^{-+}$ & ($\psibar \gamma_5 \psi$ , $\psibar \gamma_5\gamma_0 \psi$) & ($\pi$) \\
$1^{++}$ & ($\psibar \gamma_5\gamma_i \psi$) & ($a_1$) \\
$1^{+-}$ & ($\psibar \gamma_5\gamma_0\gamma_i \psi$) & ($b_1$) \\
$1^{--}$ & ($\psibar \gamma_i \psi$ , $\psibar \gamma_i\gamma_0 \psi$) & ($\rho$) \\
\end{tabular}
\vskip0.05in
We can also give the quark and antiquark a relative orbital angular
momentum.
This may be useful, because in the nonrelativistic quark model the $a_1$
($1^{++}$),
and hence the $0^{+-}$ and $0^{--}$ exotics constructed below,
are P wave states. 
The operator $\diffpartial_i = \rightpartial_i - \leftpartial_i$
inserted in the quark bilinear
brings in quantum numbers $1^{--}$,
where the negative charge conjugation
comes because C interchanges the quark and antiquark.
Thus, a P-wave operator with $a_1$ quantum numbers, $1^{++}$, is
$ \epsilon_{ijk} \psibar \gamma_j \diffpartial_k \psi $.
This operator may be advantageous because it couples the ``large''
components of the quark spinor to the large components of the antiquark
spinor.
For $F_{\mu\nu}$ we use a ``pointlike'' construction, illustrated here.
Each open loop represents
the product of the links, minus the adjoint of the product.
Each of these links may actually be a ``smeared'' link, as illustrated
on the right side of the figure.

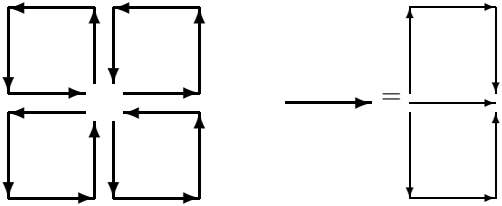
\setlength{\unitlength}{0.5in}
\begin{picture}(6.0,2.3)(-1.5,-1.15)	

\thicklines
\put(0.2,0.1){\vector(1,0){0.8}}
\put(1.0,0.1){\vector(0,1){0.9}}
\put(1.0,1.0){\vector(-1,0){0.9}}
\put(0.1,1.0){\vector(0,-1){0.8}}

\put(-0.1,0.2){\vector(0,1){0.8}}
\put(-0.1,1.0){\vector(-1,0){0.9}}
\put(-1.0,1.0){\vector(0,-1){0.9}}
\put(-1.0,0.1){\vector(1,0){0.8}}

\put(-0.2,-0.1){\vector(-1,0){0.8}}
\put(-1.0,-0.1){\vector(0,-1){0.9}}
\put(-1.0,-1.0){\vector(1,0){0.9}}
\put(-0.1,-1.0){\vector(0,1){0.8}}

\put(0.1,-0.2){\vector(0,-1){0.8}}
\put(0.1,-1.0){\vector(1,0){0.9}}
\put(1.0,-1.0){\vector(0,1){0.9}}
\put(1.0,-0.1){\vector(-1,0){0.8}}

\put(1.9,0.0){\vector(1,0){0.9}}
\put(2.9,0.0){=}
\thinlines
\put(3.2,0.0){\vector(1,0){0.9}}
\put(3.2,0.1){\vector(0,1){0.9}}
\put(3.2,1.0){\vector(1,0){0.9}}
\put(4.1,1.0){\vector(0,-1){0.9}}
\put(3.2,-0.1){\vector(0,-1){0.9}}
\put(3.2,-1.0){\vector(1,0){0.9}}
\put(4.1,-1.0){\vector(0,1){0.9}}

\end{picture}

To make a color singlet including the octet $F_{\mu\nu}$ we take the
color structure $\psibar^a F^{ab} \psi^b$,
where $a$ and $b$ are color indices.
We have computed propagators for the following hybrid operators:

\psfigure 3.0in 0.0in {NONEXOTIC_FIG} {props.nonexotic.ps} {
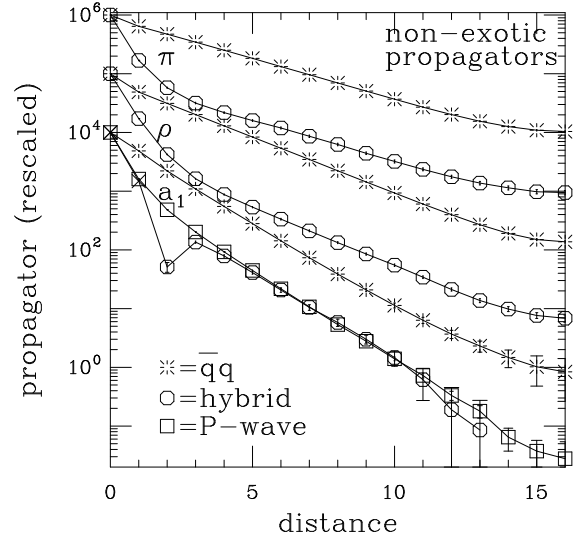
Propagators with conventional quantum numbers $0^{-+}= \pi$,
$1^{--}=\rho$ and $1^{++}=a_1$.
Propagators are scaled so that all propagators
for the same quantum numbers begin at the same point.
Hybrid propagators with these quantum numbers show the same
mass (slope) as the quark-antiquark operators, indicating mixing between
the hybrid and $\bar q q$ components.  These propagators are for
$\kappa=0.1585$.
}

\psfigure 3.0in 0.0in {EXOTIC_FIG} {props.exotic.ps} {
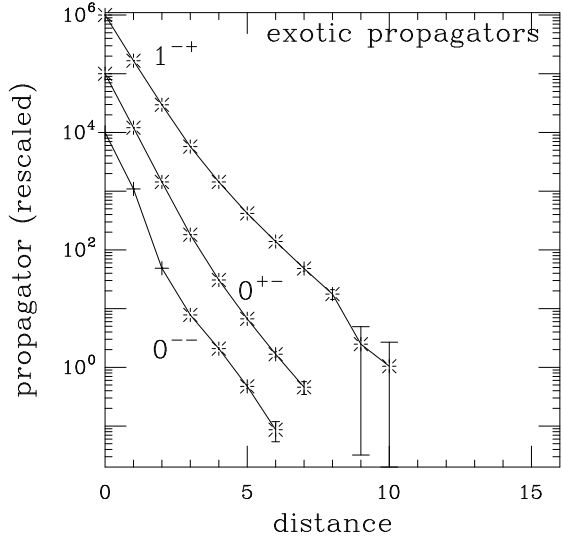
Hybrid propagators with exotic quantum numbers.  These propagators
use a twice-smeared $F_{\mu\nu}$ at $\kappa=0.1585$.
Again, propagators have
been rescaled to arbitrary values at distance zero.  The plus signs in
the $0^{--}$ propagator indicate a change of sign.
The scale is the same as Fig.~\protect\ref{NONEXOTIC_FIG} to
facilitate comparison.
}

\hrulefill\vskip 0.0in
$0^{-+}$: ({\it Mixes with pion})\\
Take $1^{--}$ quark bilinear and $\vec B$:\\
\BE 
0^{-+} = \epsilon_{ijk} \psibar^a \gamma_i \psi^b F_{jk}^{ab}
\EE

\hrulefill\vskip 0.0in
$1^{--}$: ({\it Mixes with rho})\\
Take $0^{-+}$ quark bilinear and $\vec B$:\\
\BE
1^{--} = \epsilon_{ijk} \psibar^a \gamma_5 \psi^b F_{jk}^{ab}
\EE

\hrulefill\vskip 0.0in
$1^{++}$: ({\it Mixes with $a_1$})\\
Take $1^{--}$ quark bilinear and $\vec E$:\\
\BE
1^{++} = \epsilon_{ijk} \psibar^a \gamma_j \psi^b F_{0k}^{ab}
\EE

\hrulefill\vskip0.0in
$0^{+-}$: ({\it Exotic})\\
Take $1^{++}$ quark bilinear and $\vec B$:\\
S wave:
\BE
0^{+-} = \psibar^a \gamma_5\gamma_i \psi^b B_i^{ab} 
= \psibar^a \gamma_5\gamma_i \psi^b \epsilon_{ijk}F_{jk}^{ab}
\EE
P wave:
\BE
0^{+-} = \epsilon_{ijk} \psibar^b \gamma_j \diffpartial_k \psi^a
B_i^{ab} 
= \psibar^b \gamma_j \diffpartial_k \psi^a F_{jk}^{ab}
\EE

\hrulefill\vskip 0.0in
$0^{--}$: ({\it Exotic})\\
Take $1^{++}$ quark bilinear and $\vec E$:\\
S wave:
\BE
0^{--} = \psibar^a \gamma_5\gamma_i \psi^b E_i^{ab}
= \psibar^a \gamma_5\gamma_i \psi^b F_{i0}^{ab}
\EE
P wave:
\BE
0^{--} = \epsilon_{ijk} \psibar^b \gamma_j \diffpartial_k \psi^a
E_i^{ab}
= \epsilon_{ijk} \psibar^b \gamma_j \diffpartial_k \psi^a F_{0i}^{ab}
\EE

\hrulefill\vskip 0.0in
$1^{-+}$: ({\it Exotic})\\
take a $1^{--}$ quark operator and $\vec B$:\\
$1^{--} \otimes 1^{+-} = 0^{-+} \oplus {\bf 1^{-+}} \oplus 2^{-+}$
\BE
1^{-+} = \eps_{ijk} \psibar^a \gamma_j \psi^b  B_k^{ab}
= 2 \psibar^a \gamma_j\psi^b F_{ji}^{ab}
\EE

\vskip-0.1in
\section{Simulations}

\psfigure 3.0in 0.0in {EMASS_FIG} {emass.1mp.ps} {
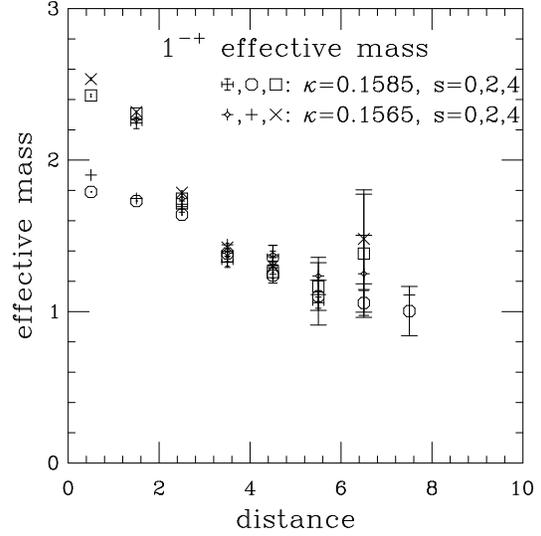
Effective mass for the $1^{-+}$ propagator.
``s'' is the number of iterations of smearing in computing $F_{\mu\nu}$.
}

\psoddfigure 3.0in 2.9in 0.0in {FIT_FIG} {1mp_mass.ps} {
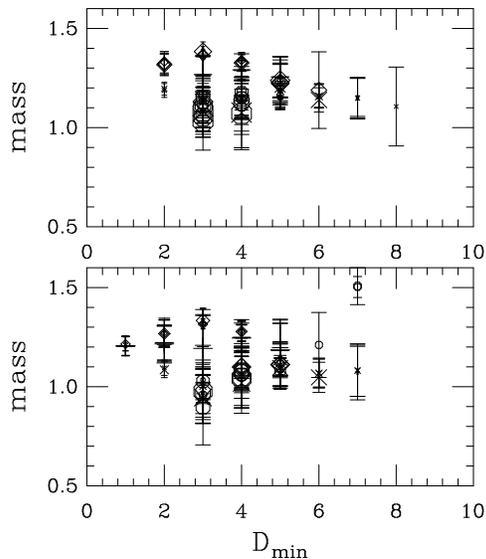
Mass fits for the $1^{-+}$ propagator.  We restrict to fits with
confidence level greater than 0.1 and error less than 0.2.
$\kappa=0.1565$ is on the top and $\kappa=0.1585$ on the bottom.
Diamonds are for an unsmeared source, crosses for the smear=2 source,
plusses for the smear=4 source, and octagons for the simultaneous 
two-propagator fits.
Both one-mass and two-mass fits are included.
The symbol size is proportional to the confidence level of the fit.
}

Propagators were computed on a subset of the HEMCGC two flavor lattices.
The lattice size is $16^3\times 32$, the sea quarks are Kogut-Susskind
quarks with $am_q=0.01$, and the gauge coupling is $6/g^2=5.6$.
We use Wilson valence quarks in computing propagators,
with $\kappa=0.1565$ and $0.1585$.
Many quantities have already been calculated on these lattices,
including the conventional meson spectrum at these
$\kappa$\cite{HEMCGC_COLD_WILSON}.  Estimates for the lattice spacing
range from 1.8 to 2.4 GeV, depending on what quantity is chosen as
the standard.

\vskip0.05in\leftline{Conventional meson masses}\nobreak
\begin{tabular}{llll}
$\kappa$ &$am_\pi$ &$am_\rho$ &$am_{a_1}$ \\
0.1565	& 0.447(1) & 0.526(3) & 0.820(12) \\
0.1585 & 0.331(1) & 0.442(4) & 0.699(8) \\
\end{tabular}
\vskip 0.05in
We used wall sources and point sinks, summed over space to get zero momentum,
with four source slices per lattice.
The exotic propagators shown here use P-wave sources for $0^{+-}$ and $0^{--}$,
and an S-wave source for the $1^{-+}$.

We show hybrid propagators for conventional and exotic quantum
numbers in Figs.~\ref{NONEXOTIC_FIG} and \ref{EXOTIC_FIG}.
Of the three exotics studied, the $1^{-+}$ is the
lightest.  Effective masses for the $1^{-+}$ are in Fig.~\ref{EMASS_FIG}.
Although a better plateau in the effective mass is clearly desirable,
we proceed to fit the propagator to an exponential over various ranges.
The results are summarized in Fig.~\ref{FIT_FIG}.  Of course, all our
source operators (different smearings) should give the same masses,
with different amplitudes depending on the overlap of the operator with
the various states.  To get mass estimates from shorter distances, we
may try fitting two or more propagators simultaneously, constraining
them to have the same masses but allowing different 
amplitudes\cite{MULTISOURCE}.
Fig.~\ref{FIT_FIG} also contains two source fits, with two masses and
four amplitudes, using smearing levels
0 and 2 simultaneously, or 2 and 4.
The mass fits come out around $am_{1^{-+}} \approx 2 am_{1^{--}}$.
If the valence quarks had the masses of the physical light quarks, this
would be $\approx 2 am_\rho$, but in fact the valence quarks are closer
to the strange quark in mass.

\section{Conclusions}

The study of hybrid mesons on the lattice is in its infancy.
These calculations combine the worst features of glueball
and quark-antiquark spectrum calculations --- operators with large
fluctuations, with time consuming propagator inversions.
The situation is like the early glueball calculations, 
with only short propagators available.
We need more work to develop good
operators, more statistics, and studies of dependence on quark mass and
lattice spacing. 
Our results show that these calculations are doable, and we are
optimistic about the ability of lattice calculations to find the
spectrum of hybrid particles.

Calculations were done on
the Intel Paragons at SDSC
and Indiana University, the Alpha Cluster at
PSC and the SP2 at CTC. 
This research was supported by the United States NSF and DOE.

\end{document}